\documentclass{article}
\pdfoutput=1 

\usepackage{array}
\usepackage{pgfplots}
\usepackage{graphicx}
\usepackage{natbib}
\usepackage[margin=2.4cm]{geometry}

\begin{document}

\title{Multi-dimensional Urban Network Percolation}

\author{Juste RAIMBAULT$^{1,2,3,\ast}$\\
$^1$ UPS CNRS 3611 ISC-PIF, France\\
$^2$ CASA, UCL, UK\\
$^3$ UMR CNRS 8504 G{\'e}ographie-cit{\'e}s\\\medskip
$^{\ast}$ \texttt{juste.raimbault@polytechnique.edu}
}

\date{}

\maketitle

\abstract{Network percolation has recently been proposed as a method to characterize the global structure of an urban system form the bottom-up. This paper proposes to extend urban network percolation in a multi-dimensional way, to take into account both urban form (spatial distribution of population) and urban functions (here as properties of transportation networks). The method is applied to the European urban system to reconstruct endogenous urban regions. The variable parametrization allows to consider patterns of optimization for two stylized contradictory sustainability indicators (economic performance and greenhouse gases emissions). This suggests a customizable spatial design of policies to develop sustainable territories.
}

\section{Introduction}

The structure of road networks can be used as a proxy to understand its past growth dynamics, but also has a significant impact on the future sustainability of territories it irrigates. Diverse methods to characterize the structure of spatial networks, and more particularly road networks, have been developed in that context, including classical network indicators such as centralities \citep{crucitti2006centrality} but also more elaborated constructions capturing more realistic processes in terms of street network use \citep{lagesse2015spatial}. These study are by essence interdisciplinary, or at least imply complementary viewpoints from disciplines as diverse as architecture with space syntax \citep{hillier1976space}, physics with the study of spatial networks \citep{barthelemy2011spatial}, or social science disciplines concerned with space such as geography \citep{ducruet2014spatial}.

A method to characterize topologies of such urban spatial networks is network percolation, initially applied to road networks by \cite{arcaute2016cities}. Percolation in physics can be understood in a broad sense as processes related to the progressive occupation or connection of nodes of a network, and is generally associated to a phase transition with the emergence of a giant cluster at a given connection probability \citep{stauffer2014introduction}. Important applications include the quantification of network robustness \citep{callaway2000network} or the modeling of epidemic spreading \citep{newman1999scaling}. Such approaches have been applied to urban systems not only for the study of networks. \cite{makse1998modeling} model urban growth with a local percolation model for site occupancy. \cite{arcaute2016cities} focus on the analysis of street networks and extract endogenous urban regions for UK which correlate with socio-economic properties, and provide a definition of urban areas which highly correlates with land-cover data. \cite{piovani2017urban} apply road network percolation at the mesoscopic scale of London metropolitan area, in relation with a retail location model. In spatial statistics, this method can be used to characterize the spatial morphology of point patterns \citep{huynh2018characterisation}.

Existing heuristics however generally focus on a single dimension or property of the urban system. However, such systems are known to be multidimensional, for example through the superposition of the morphological dimension of networks, and the functional properties of the urban environment \citep{burger2012form}. The link between urban form and function remains in particular an open question \citep{batty1994fractal}, but more generally the inclusion of multiple dimension in urban analysis is still a research direction to be investigated, as in the case of agent-based models for example \citep{perez2016agent}. This paper addresses such a gap in the case of urban network percolation, by introducing a multi-dimensional percolation heuristic. The method allows to combine different dimensions of the urban system, the same way that \cite{cottineau2018defining} combines population density and commuting flows to produce multiple definitions of urban areas.

Beside these methodological issues, applied tools are needed to quantify the sustainability of recently emerged urban forms. In particular, according to \cite{lenechet2017peupler}, the most recent transition of human settlement systems (in the sense of \cite{sanders2017peupler}, i.e. a change in the dynamical regime ruling the evolution of the spatial structure of settlements) is the emergence of mega-city regions. These have been defined by \cite{hall2006polycentric} as polycentric urban structures highly integrated in terms of flows. The transition imply complex processes such as changes in the governance structure, and can not be associated to the stylized transition identified by \cite{louf2013modeling} in a simplistic toy urban model, and therefore has chances to imply more drivers than negative externalities of congestion only. To what extent these new urban forms are sustainable, for example in the broad sense of UN development goals \citep{komiyama2006sustainability}, remains an open question. In order to test our multi-dimensional percolation method, we propose to apply it to the endogenous characterization of urban regions, and compute stylized sustainability indicators on the constructed regions.

Our contribution relies on several points: (i) this is to the best of our knowledge the first time a multi-dimensional percolation method is applied to urban systems; (ii) we furthermore apply it on the significant spatial extent of all European Union; and (iii) we link the clusters obtained with simple sustainability measures. The rest of this paper is organized as follows: we first describe the multi-dimensional percolation heuristic, the data and variables to which it is applied, and the indicators used to characterized the sustainability of clusters produced. We then describe the results of applying this method to population and network variables for the whole European Union, focusing on the endogenous regions produced and their sustainability properties. We finally discuss possible developments and the implications of this methodology to the design of policies.

\section{Methods}

\subsection{Multi-dimensional percolation}

\begin{figure*}[ht] 
  {\includegraphics[width=\linewidth]{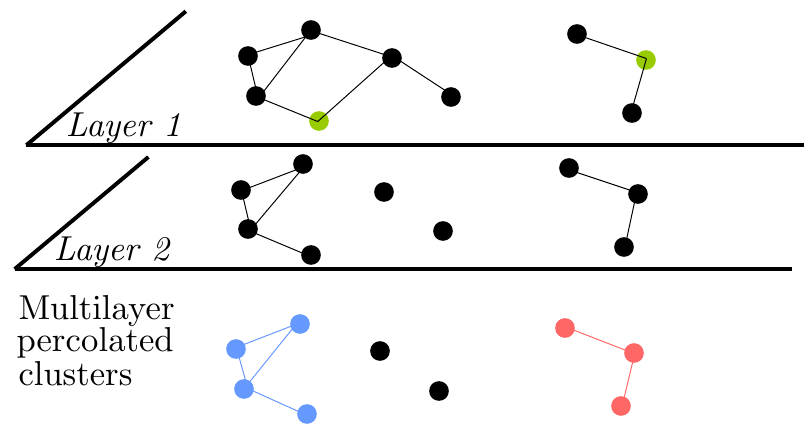}}
  \centering
  \caption{Schematic representation of the multi-dimensional network percolation heuristic. We show a stylized configuration with two layers having the same nodes, and the links within each layer are created following the percolation radius $r_0$ and the thresholds for the layer variables. The final clusters are the superposition of these. The green points give examples of starting points in the case of a propagation heuristic.\label{fig:method}}
\end{figure*}

Percolation processes in multilayer networks have been proposed as an extension within simple networks \citep{boccaletti2014structure}. A generalization of epidemic spreading can for example be achieved using this framework \citep{son2012percolation}. In the case of multilayer networks sharing the same nodes for all layers, often called multiplex networks, bond percolation has also been studied \citep{hackett2016bond}.

In the case of our heuristic, bond percolation is operated between two cells given a distance threshold, and furthermore with a threshold parameter for each layer assuming a node function within each layer. The distance-based connection is similar to generative processes for random euclidian networks \citep{penrose1999k}.

More formally, let assume a set of nodes $V = v_i$ common to all layers, and layers edges $E_j$ taken as empty at the initial state of the algorithm. Each node has a value of the considered variables associated to each layer, written $v_{ij}$. For each layer, a link $e_{kl} \in E_j$ is created if $d(v_k,v_l) < r_0$ where $d$ is the distance between the nodes (which can be any distance) and $r_0$ the percolation radius, and if $v_{kj} > \theta_j$ and $v_{lj} > \theta_j$ where $\theta_j$ is the threshold for layer $j$. The final percolated network edges $E$ is composed by links contained within all layers simultaneously. The multi-dimensional percolation clusters are then the connected components of this network $(V,E)$. The parameters implied in this heuristic are the percolation radius $r_0$ and the percolation thresholds $\theta_j$ for each layer $j$, allowing a flexible application through parametrization.

Note that we do not call our method ``multi-layer percolation'', as it is not strictly multi-layer since nodes are common. The term of multi-dimensional percolation is more suited to the use of multiple variables and thresholds. The method works with an arbitrary number of layers. See Fig.~\ref{fig:method} for a schematic representation of the method. It can be implemented with a propagation heuristic or directly working on adjacency matrices. The rationale behind the conjunction of the thresholding of each layer variable and the distance thresholding relies on the idea that two points will interact is they are close enough, but also if they have a strong enough intensity of the activity or dimension captured by each layer, simultaneously for all layers considered. This recalls Tobler's first law of geography \citep{tobler2004first} in a multi-dimensional way.

\subsection{Empirical data}

We apply the heuristic to urban morphology and road network topology measures in Europe. The idea to combine urban form with network topology measures relies on the capture of the link between urban form and function as already mentioned, urban functions being assumed as distributed by transportation networks \citep{raimbault2018caracterisation}.

More precisely, a grid of population density morphology indicators and road network topology indicators has been computed on spatial moving windows of width 50km for all European Union by \cite{raimbault2018urban}, with an offset resolution of 5km. We use this data to construct a two layers abstract network: a layer which variable is given by population density, and a second layer which variable is given by a network variable. Nodes are the center of cells (thus disposed in space on a grid of step 5km). We test the variable characterizing the second layer among the following characteristics of the road network within the corresponding window: number of edges $N_E$, number of vertices $N_V$, cyclomatic number $\mu$ and euclidian efficiency $v_0$. These measures capture functional properties especially for the two last.

The percolation on such an abstract network is a necessary condition in our case to link the different dimensions considered, namely population distribution and local road network properties. We have therefore two levels of networks in our approach, namely the physical road network which local properties are taken here as input, and the abstract two layer network on which we do the percolation. We will in the following write $\theta_P$ for the threshold parameter of the population layer, and $\theta_N$ for the threshold parameter of the network layer. In practice, these parameters will be given in the following as quantile level of the corresponding variable, for an easier interpretation and conception of experience plans. The name of the variable considered will be written $v_N$.

\subsection{Sustainability indicators}

As already detailed, recent forms of urbanization, in particular integrated mega-city regions, may imply different patterns of economic and transportation flows and thus exhibit various performances regarding different indicators of sustainability. We propose to use the endogenous definition of regional urban systems produced by the percolation algorithm to evaluate their sustainability, in terms of conflicting objectives of economic integration and greenhouse gases emissions. The definition of sustainability, or sustainable development, is by essence multi-dimensional \citep{viguie2012trade}. Its characterization as quantitative indicators is even more subject to numerous degrees of freedom. We work here with two stylized indicators for two conflicting dimensions, as a proof-of-concept.

We use the EDGAR database \citep{janssens2017edgar} (version 4.3.2) for local grid estimates of greenhouse gases emissions. We use the latest year available, namely 2012. As its resolution is much smaller than our indicator grid, we aggregate the emissions on the closer indicator point for each cell of the emission database. Since according to \cite{lashof1990relative} most of the greenhouse effect is caused by $\textrm{CO}_2$, and as in terms of emissions in the database we find that it represents $98.2\%$ in mass proportion of all gases, we only consider it.

Applying a gravity model to each region, we estimate abstract transportation flows within each and use these to extrapolate emissions from the actual local emission from the Edgar database, and economic activities with a scaling law of population.  More precisely, following \cite{raimbault2018indirect}, a potential flow between two points $i$ and $j$ can be estimated with the following expression 

\begin{equation}
\phi_{ij}^{(k)} = \left(\frac{v^{(k)}_i v^{(k)}_j}{(\sum_l v_l)^2}\right)^\gamma \cdot \exp\left(\frac{-d_{ij}}{d_0}\right)
\end{equation}

where $v^{(k)}_i$ are either population or effective local GHG emissions (indexed by $k = 1,2$ respectively), $d_{ij}$ the distance between the two points, $d_0$ a distance decay parameter, and $\gamma$ a scaling parameter. Indeed, the economic activity follows relatively well scaling laws of populations \citep{bettencourt2007growth}, the exponent being dependant on the activity and the definition of areas on which it is estimated~\citep{cottineau2017diverse}.

The sum of all flows of points within the geographical span of the cluster (that we approximate as the convex Hull envelope of its points), allows us to approximate the cumulated potential emissions and economic activity. Writing clusters $K_c$ as this set of points, we define the total economic flow by $E_c = \sum_{i,j \in K_c} \phi_{ij}^{(1)}$ and the total emissions due to flows by $G_c = \sum_{i,j \in C_c} \phi_{ij}^{(2)}$. This allows to define a relative economic inefficiency by $e_c = 1 - \frac{\max_c E_c - E_c}{\max_c E_c - \min_c E_c}$ and relative potential emissions by $g_c = \frac{\max_c G_c - G_c}{\max_c G_c - \min_c G_c}$. Both indicators should be minimized for sustainability. Normalized indicators $\tilde{e}_c,\tilde{g}_c$ are defined in a similar way, but the extrema being computed on all other possible urban configurations with the same $\gamma,d_0$ values.

Using these potential flows follows the logic of \cite{arbabi2019development} which shows a need for improved intra-city-region mobility in England and Wales. Considering the regions as entities in which such transportation development policies can more easily been developed, we look at the sustainability of different possible regions if these potential flows were realized. Varying the parameters $\gamma$ and $d_0$ allows to control for the economic activity considered (high $\gamma$ values correspond to high added-value activities) and the span of interactions through $d_0$.

\section{Results}

\subsection{Implementation}

In practice, the analysis is implemented using R and the igraph package. Source code, data and results are available on the open git repository of the project at \texttt{https://github.com/JusteRaimbault/UrbanMorphology}. The network is constructed by superposing the population density layer with the network layer, starting from the 5km resolution spatial fields for morphological and network indicators. This network is filtered with the threshold parameters for each layer and with the radius parameter. Connected components yield the clusters that we interpret as endogenous regions.

We recall that the euclidian performance of the network \citep{banos2012towards} is in our case $<d_e/d_n>$ where the average is taken on all origin-destination pairs in the network, $d_e$ is the euclidian distance and $d_n$ the network distance. Thus, it indeed increases with network performance, in consistence with the use done here through thresholding.

\begin{figure*}[ht] 
  {\includegraphics[width=0.49\linewidth]{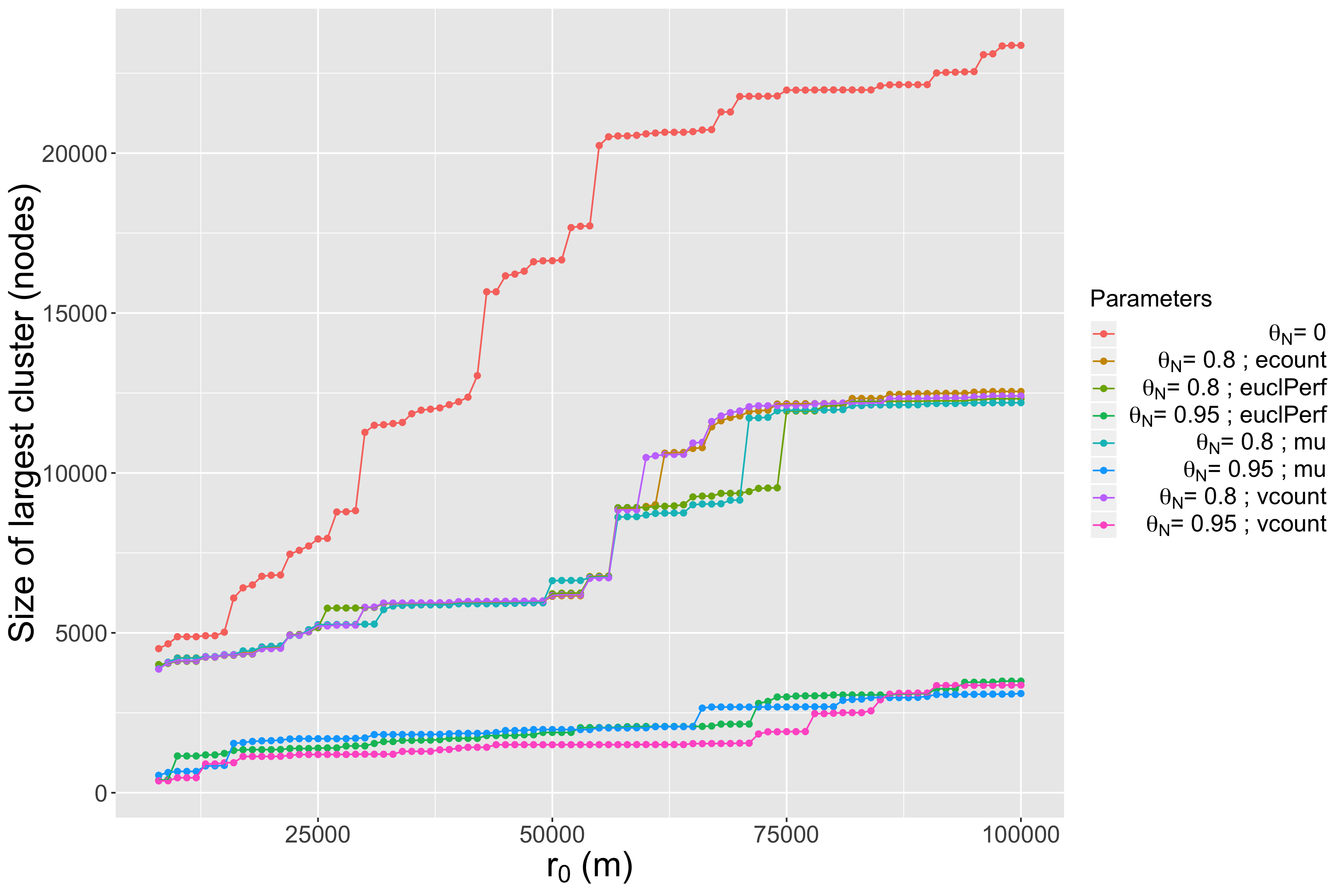}}
  {\includegraphics[width=0.49\linewidth]{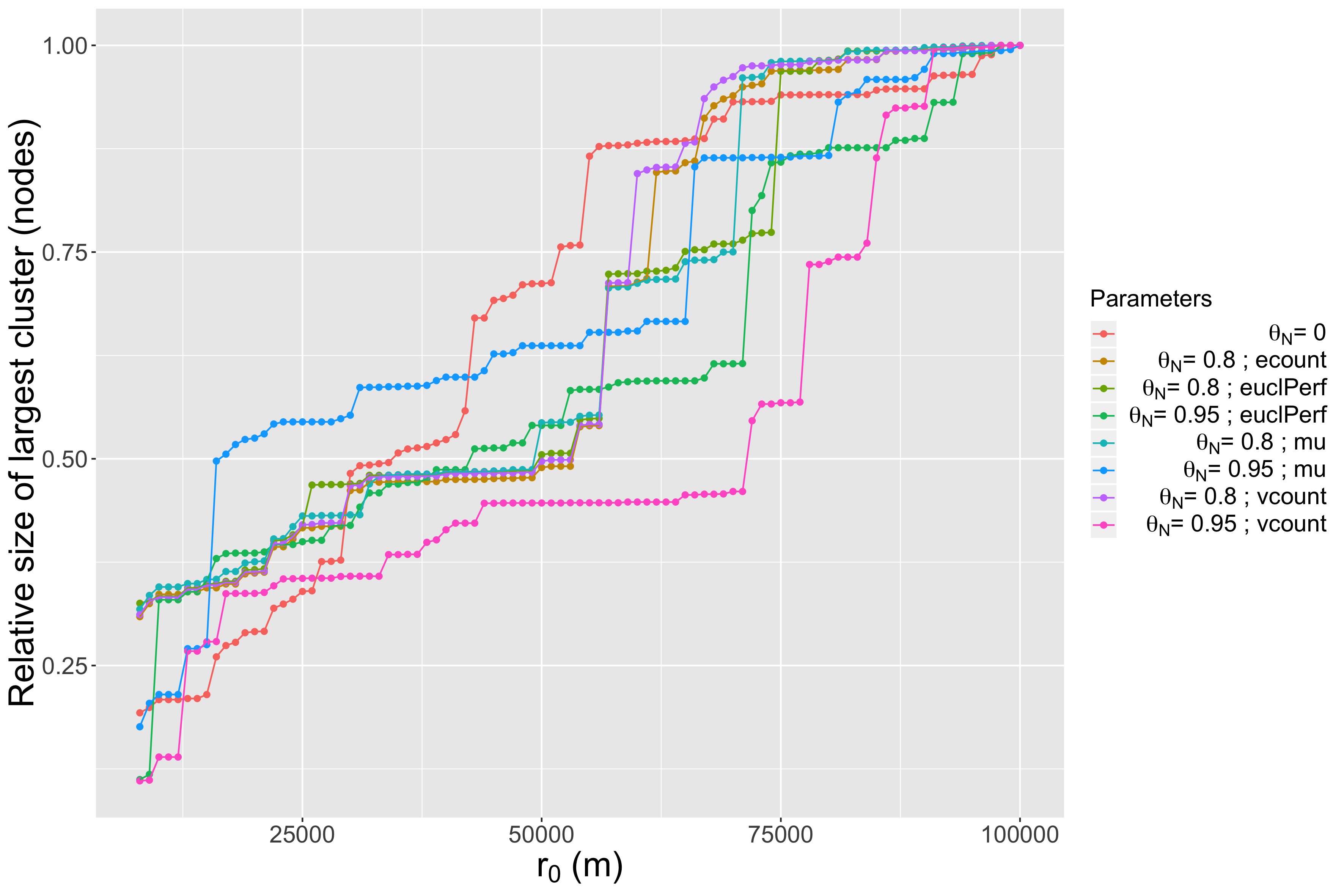}}
  \centering
  \caption{Percolation transition. On the left, we plot the size of the largest cluster in each configuration in terms of nodes, as a function of the percolation radius $r_0$. Color gives the other percolation parameters. On the right, the plot is similar but with the size relative to the size of the largest cluster obtained with the maximal radius in each configuration.\label{fig:percolation}}
\end{figure*}

\begin{figure*}[ht] 
  {\includegraphics[width=\linewidth]{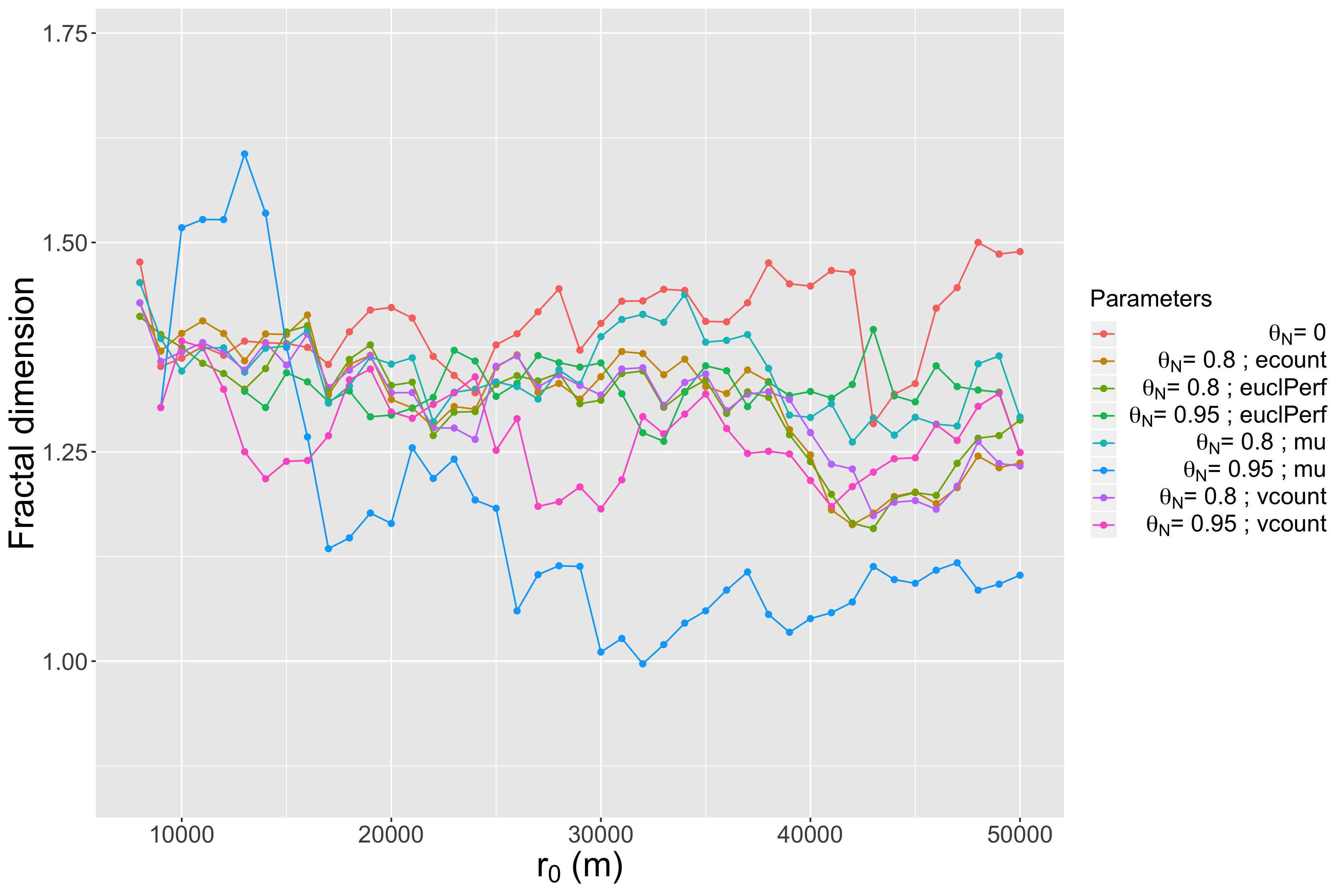}}
  \centering
  \caption{Fractal dimension. We plot for each parametrization given by the curve color the evolution of the fractal dimension $\alpha$ as a function of $r_0$. Standard errors are not plotted for readability.\label{fig:fractaldim}}
\end{figure*}

\subsection{Percolation transition and fractal dimension}

In its application to road networks by \cite{arcaute2016cities}, the structure of the national urban system for UK is captured by studying the percolation transition, i.e. the variation of the size of the largest cluster as a function of the percolation radius. As this signature is tightly linked to historical, cultural and geographical conditions, the application to different urban systems should yield different results. We study here this property, for different threshold parameter values. We make the radius vary betweem 8km and 100km with a one km step, have a fixed population threshold $\theta_P = 0.85$, test all network variables, and three network thresholds $\theta_N \in \{ 0 ; 0.8 ; 0.95 \}$.

The absolute and relative sizes of the largest cluster are plotted in Fig.~\ref{fig:percolation} as a function of the percolation radius. This aspect first gives methodological information on multilayer percolation. Indeed, comparing the result with $\theta_N = 0$ (single layer percolation) with positive values of $\theta_N$ shows a significantly different behavior. As expected, absolute size are much smaller, but when looking at relative sizes we observe that the abrupt steps typical to percolation transitions have different distributions across the different parametrizations. The more regular curve seems to be the standard percolation on population only, whereas at $\theta_N = 0.95$, different network variables produce either very early transitions (for $\mu$ for example) or very late (for $N_V$). Also, changing of scale compared to \cite{arcaute2016cities} gives more steps and less abrupts curves in general, confirming the integration of subsystems with different structures in our analysis and the importance of scale in such analysis. As the addition of a layer also changes drastically the results, one should stay careful when switching from a mono-dimensional percolation to a multi-dimensional percolation.


We study also the evolution of the fractal dimension of clusters as a function of $r_0$. Following \cite{arcaute2016cities}, we estimate the fractal dimension of clusters $\alpha$ with a simple OLS regression between cluster size and cluster diameter, namely $\log N_c = k + \alpha \cdot \log \delta_c$ where $N_c$ is the size of cluster $c$ and $\delta_c$ its diameter. As a negative result, that could be due to the abstract nature of our network, a clear maximum in the value of the fractal dimension can not be found. Either it is located at resolution that our method can not reached due to the minimal 5km limit imposed by the abstraction, or it does not exist when coupling dimensions. Determining which assumption is more plausible is out of the scope of this paper. We do not plot the standard error $\sigma$ of fractal dimensions for visibility purposes, but their relative value given by $\alpha / \sigma\left[\alpha\right]$ is in average 0.10 and in maximum 0.196 on all points, meaning that these estimations remain however consistent. Regarding the variability as a function of the percolation radius $r_0$, studying the difference $(\alpha - \sigma\left[\alpha\right])_M - (\alpha - \sigma\left[\alpha\right])_m$ where the first is taken at maximum for $\alpha$ and the other at minimum, shows that the configuration for $\mu$ and $\theta_N=0.95$ has a clearly significant maximum (difference at 0.38). For this coupling, the endogenous structure given by the maximum may be defined. Other configurations yield non-significant maximums.


\begin{figure}[h!] 
{\includegraphics[width=0.49\textwidth]{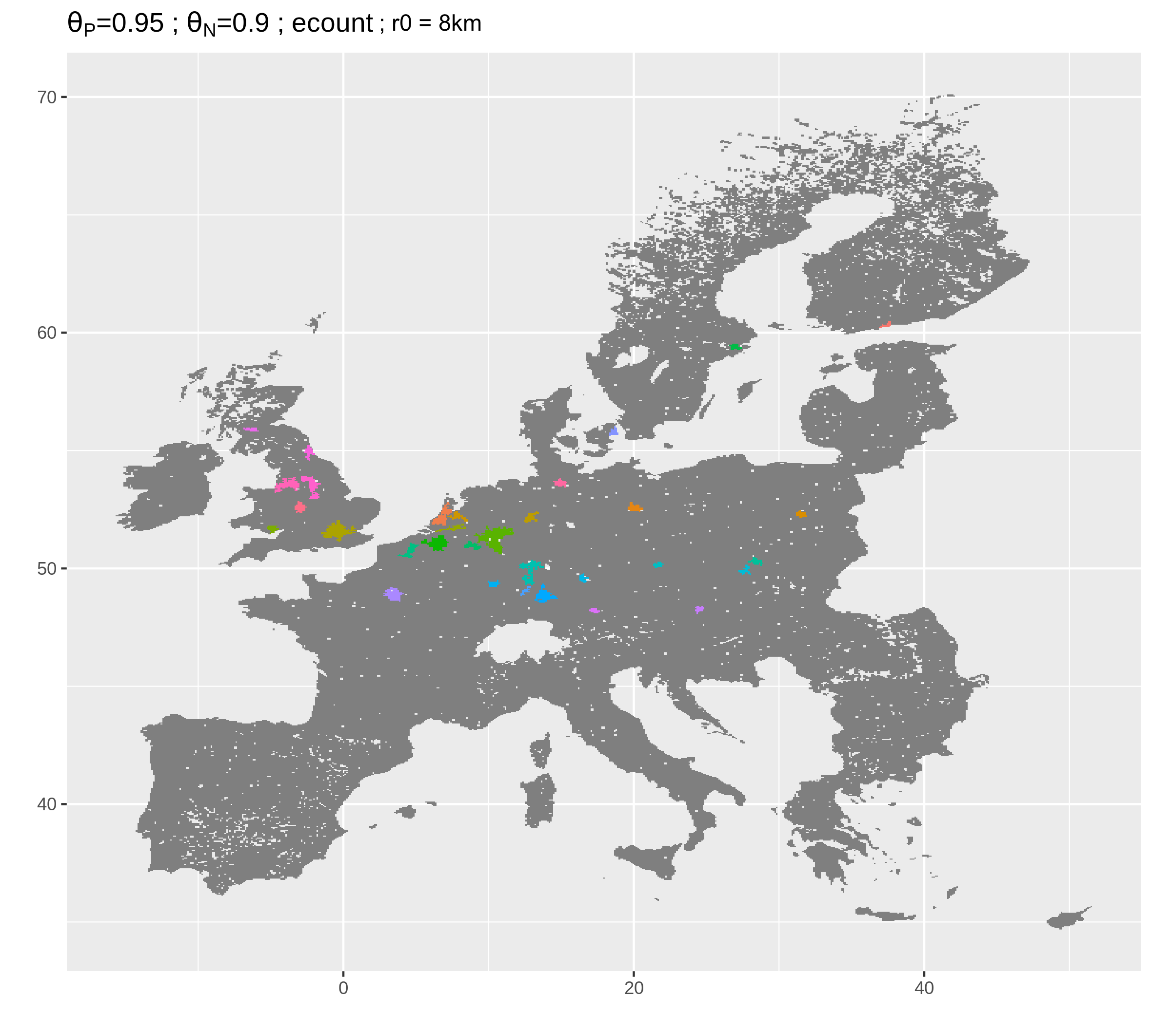}}
  {\includegraphics[width=0.49\linewidth]{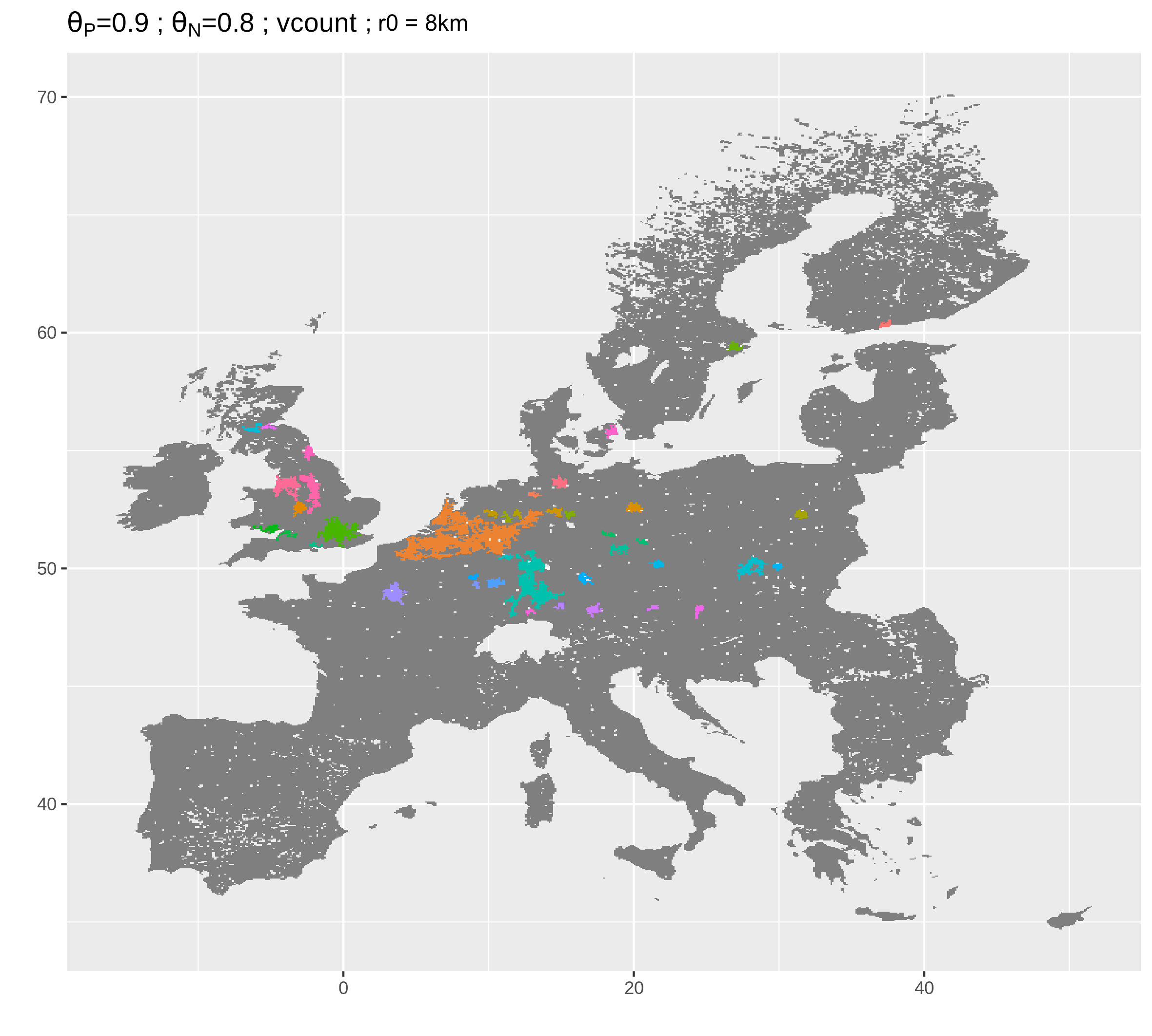}}\\
  {\includegraphics[width=0.49\linewidth]{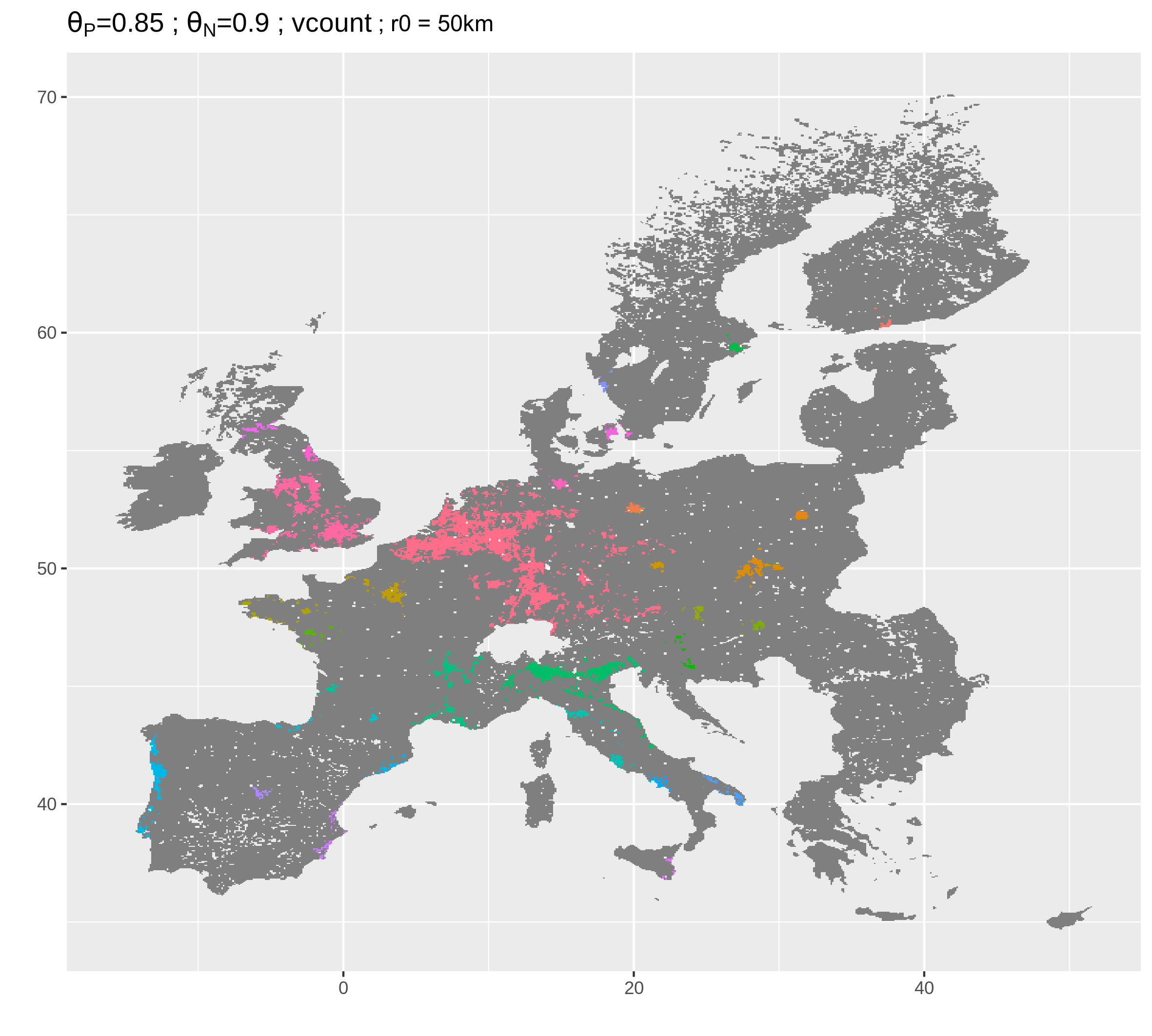}}
   {\includegraphics[width=0.49\linewidth]{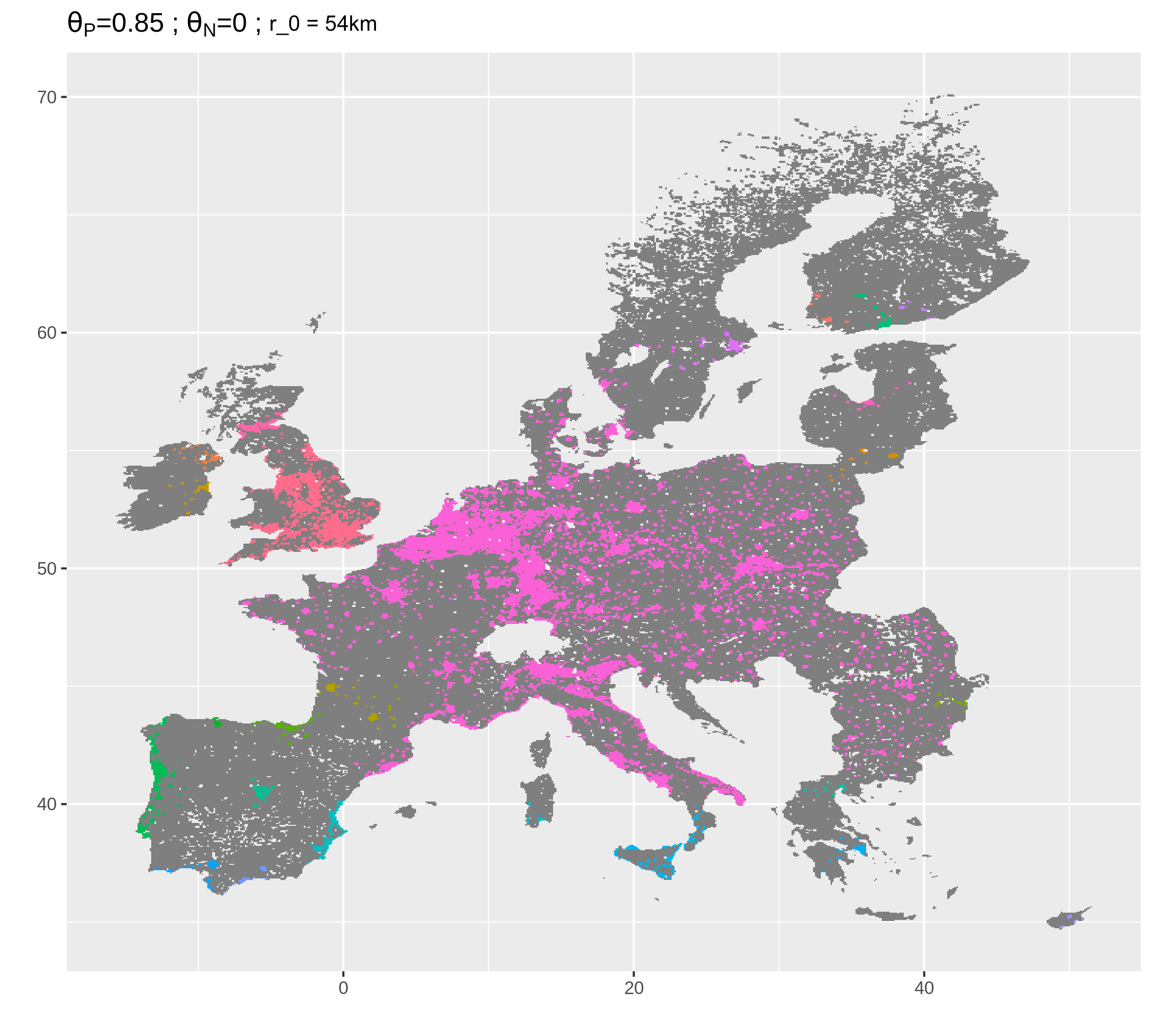}}
  \centering
  
  \caption{Examples of obtained clusters for different parameter values. In the top-right case for example ($\theta_P = 0.9$, $\theta_N = 0.8$, variable \texttt{vcount},$r_0 = 8km$), we obtain the urban regions of West midlands and London in the UK, Randstad merged with Rhein-Rhur and Rhein-Main in Germany, Paris in France, also with capital cities such as Copenhaguen, Stockholm and Helsinki. There is no cluster in South Europe in that case, due to the high population density threshold.\label{fig:exclusters}}
\end{figure}

\begin{figure}[h!] 
  {\includegraphics[width=0.7\linewidth]{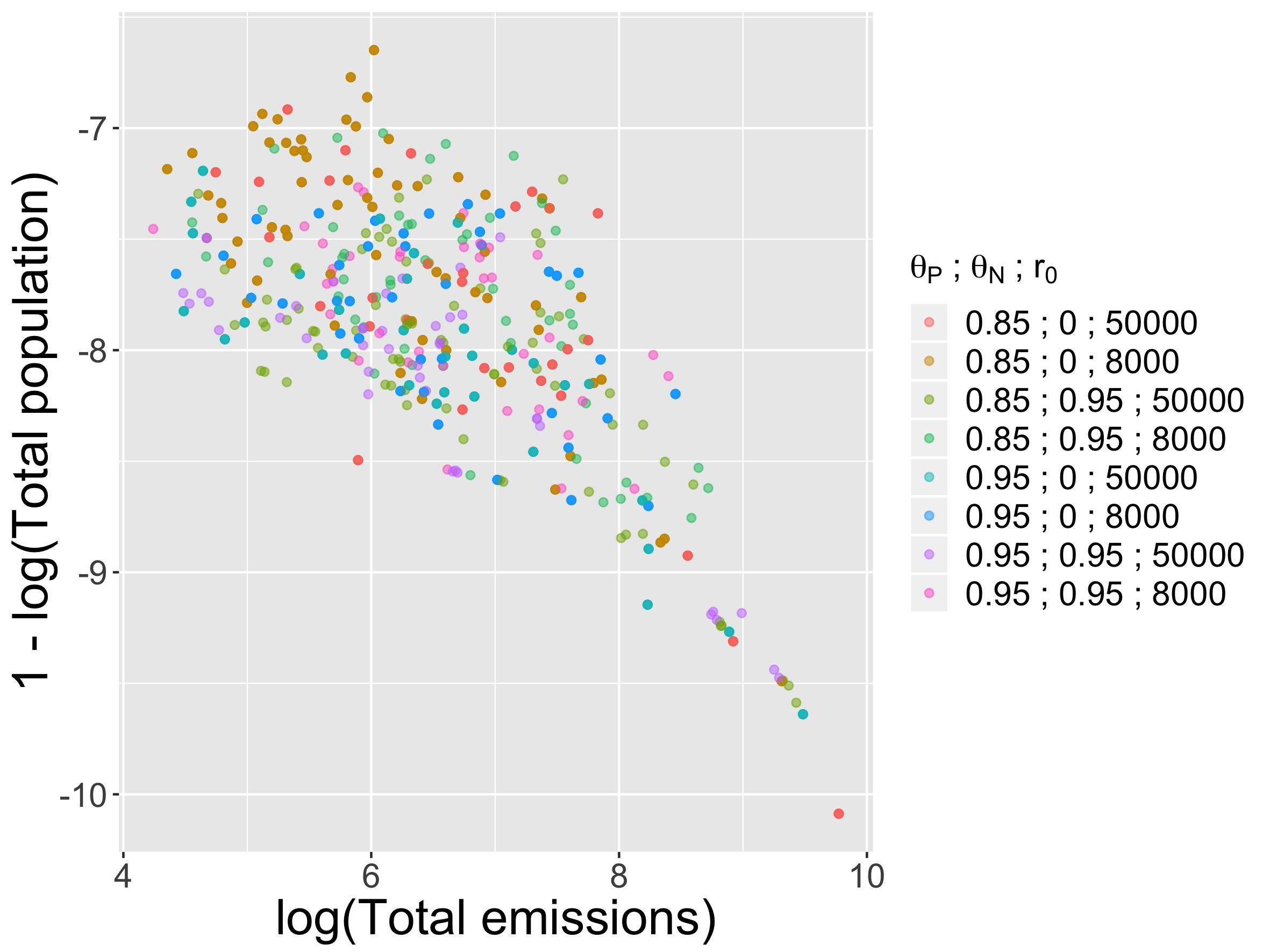}}
  \centering
  \caption{Point clouds of region-level indicators, namely population and emissions, for different parametrizations, given by the color. Each point represent an endogenous urban region.\label{fig:paretos}}
\end{figure}

\subsection{Extracting endogenous mega-city regions}

We now switch the experience plan to a full grid, for parameters $r_0$, $\theta_P$, $\theta_N$ and the network variable considered, and also make $\gamma$ and $d_0$ vary. We systematically explore the clusters obtained for 4800 parameter configurations, such that for all network variables, $\theta_P \in \{ 0.8 ; 0.9 ; 0.95 \}$, $\theta_N \in \{0 ; 0.8 ; 0.95 \}$, $r_0 \in \{ 8 ; 10 ; 15 ; 20 ; 50\}$ km, $\gamma \in \{ 0.5 ; 1 ; 1.5 ; 2\}$, and $d_0 \in \{ 0.1 ; 1 ; 10 ; 50 ; 100\}$ km.

We obtain very different endogenous morphologies for the different parametrizations. Maps reveal that some configurations resemble the actual distribution of European mega-city regions, which are functionally integrated polycentric urban areas \citep{hall2006polycentric}. These are here defined endogenously from the bottom-up and have a priori no reason to coincide with these functional regions. We show some examples in Fig.~\ref{fig:exclusters}. The first map of this figure, obtained for high population and network thresholds ($\theta_P = 0.95$ and $\theta_N = 0.9$), but a low radius $r_0 = 8$km and edge count $N_E$ as network variable, include several mega-city regions described by \citep{hall2006polycentric}, namely London metropolitan area, the Randstad in Netherland, the Rhein-Main and Rhein-Ruhr in Germany, Greater Paris in France, Brussels area in Belgium. The same parameters with $\theta_N = 0$ yield not exactly the same regions, as confirmed by the transition curves in Fig.~\ref{fig:percolation}, what means that our approach taking into account two dimensions may capture effective processes of mega-city regions, in particular by including the road network which is crucial as these regions are integrated in terms of flows. The bottom-left map show an example of large clusters emerging in UK and in the center of Europe, the South remaining largely disconnected. Finally, the last map shows the result obtained with a very high radius $r_0 = 54$km, with a giant cluster spanning most of Europe. UK is still disconnected and the transition where it connects happens at $r_0 = 55$km. This does not necessarily mean that UK should be disconnected from continental Europe, as we considered geographic distances only, hiding the high speed connection of the Channel tunnel.

The behavior of sustainability indicators for different population, network and distance thresholds yield different distributions of performances across clusters within a configuration but also between configurations. Before considering the flow-based indicators described above, we can already study basic measures such as population $P_c$ and effective emissions $EM_c$, taken as the sum within the cluster of their values at each point. We show in Fig.~\ref{fig:paretos} point clouds of $\log EM_c$ against $1 - \log P_c$ for some configurations. Indeed, regarding the population it contains, an area can be more or less efficient in terms of emissions. Seeing the population as an objective to be maximized (thus the plotted value to be minimized), we observe a Pareto front for all points (i.e. all clusters across all configurations), but also no dominating point for each configuration. Some clusters are therefore optimal compromises in the Pareto sense in each configuration, while some are dominated and thus not efficient.


\begin{figure*}[!ht] 
  {\includegraphics[width=\linewidth]{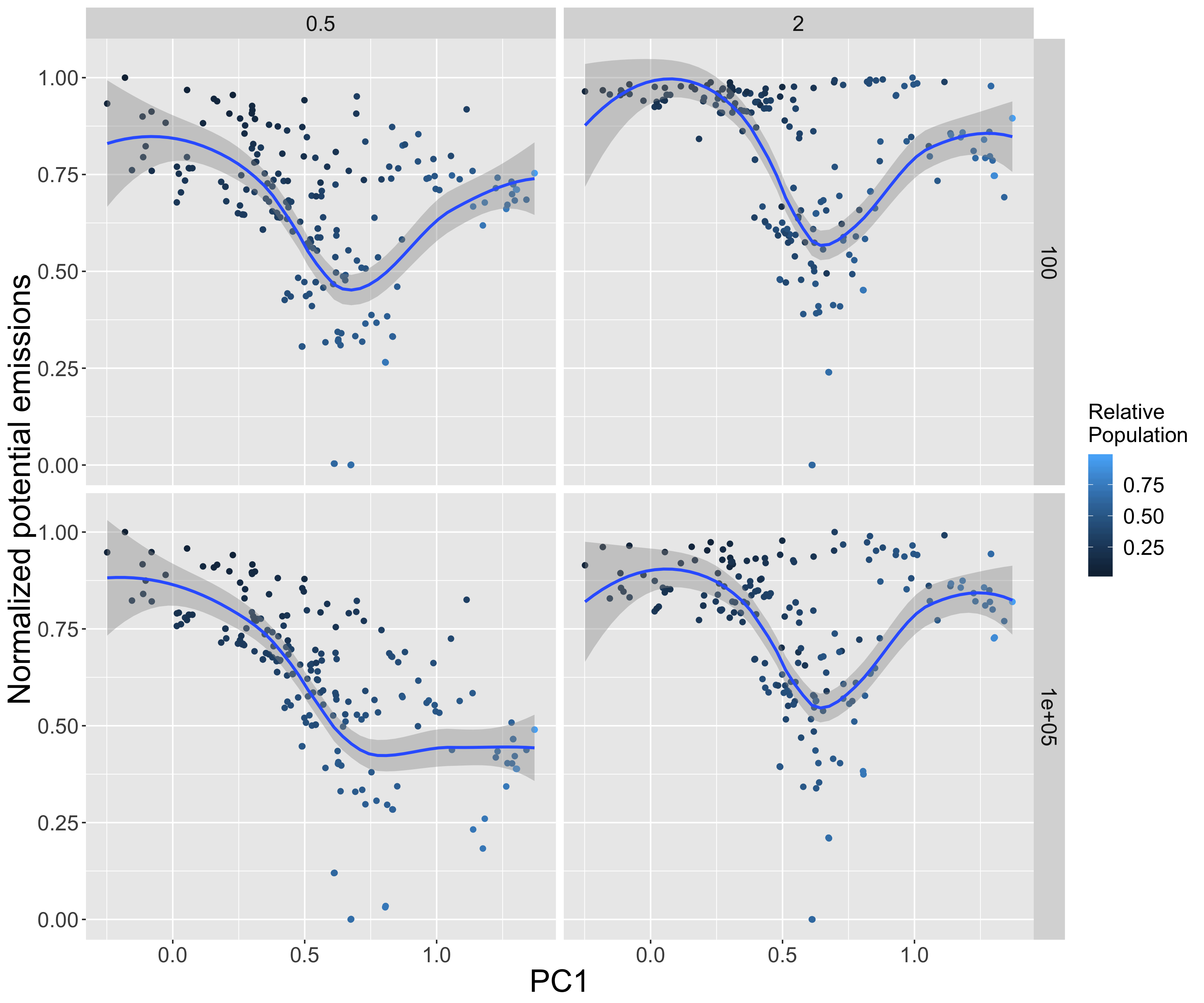}}
  \centering
  \caption{Aggregated values of normalized potential emissions $\sum_c \tilde{g}_c$, as a function of the first morphological principal component (PC1), for varying values of parameters $d_G$ (rows) and $\gamma_G$ (columns). Other intermediate values for these parameters yield similar behaviors. As PC1 is mainly linked to monocentricity, there seems to exist an optimal intermediate level of monocentricity for emissions alone. Color level give the share of population within the considered clusters in comparison to all European population.\label{fig:emissions-pc1}}
\end{figure*}

\begin{figure*}[!ht] 
  {\includegraphics[width=\linewidth]{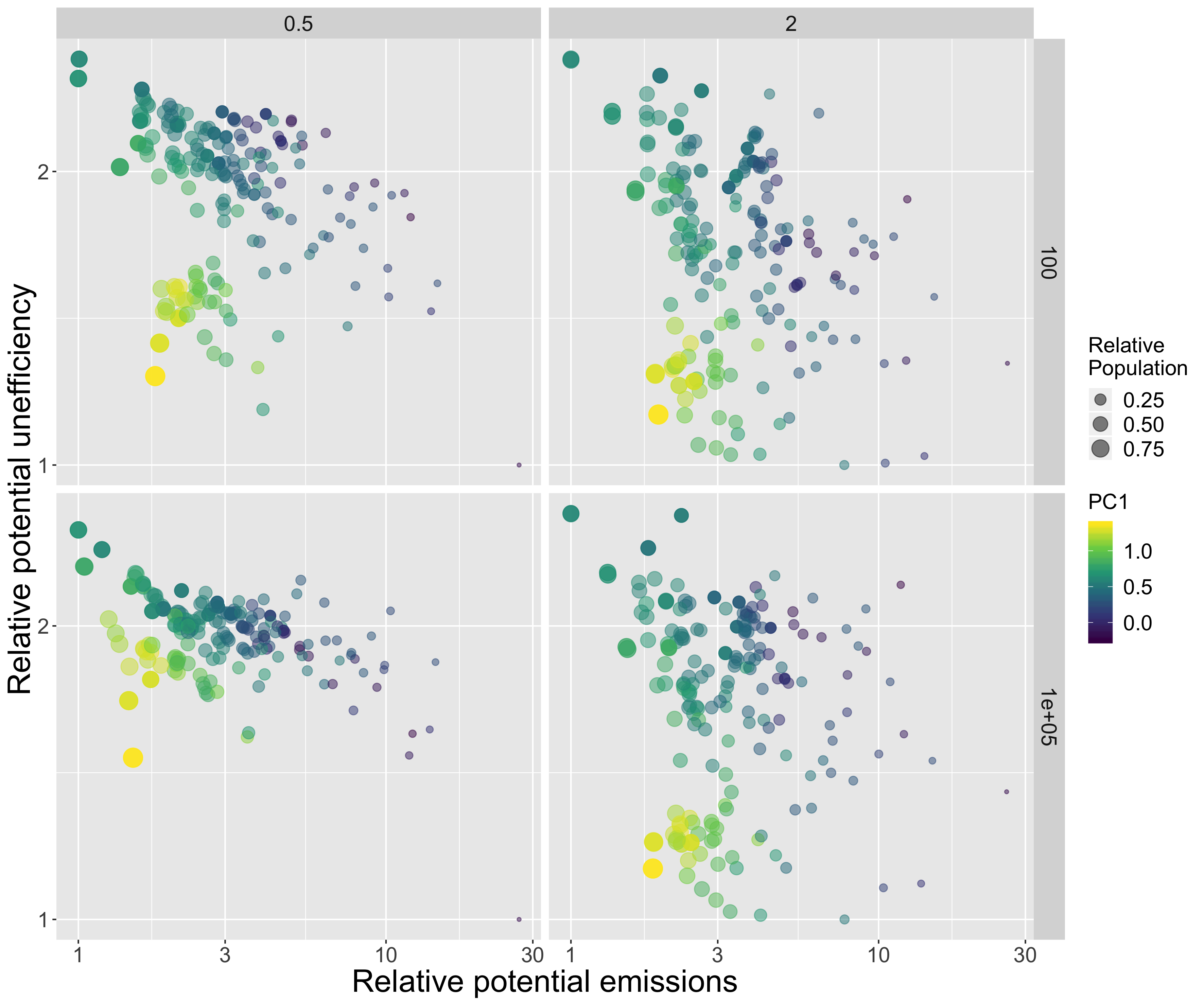}}
  \centering  
  \caption{Relative potential emissions $\sum_c g_c$ against relative potential economic unefficiency $\sum_c e_c$ (both indicators should be minimized), for varying values of $\gamma_G$ (columns) and $d_G$ (rows). Color level gives the value of PC1, whereas point size gives the share of total population contained within considered areas.\label{fig:paretos-relative}}
\end{figure*}

\subsection{Linking urban morphology and sustainability}


We now consider the sustainibility indicators, aggregated for a configuration on all clusters. For a given parametrization of endogenous city regions, one can relate them to morphological indicators for population density spatial distribution, computed by~\cite{raimbault2018calibration}, that we average on clusters. This establishes a link between urban morphology and sustainibility. A principal component analysis on considered points yield 96\% of variance with two components, and 73\% explained by the first component alone. The first component relates to a level of monocentricity ($PC1 = -0.3\cdot I + 0.54 \cdot \bar{d} + 0.51\cdot \varepsilon + 0.59 \cdot h$ where $I$ is Moran index, $\bar{d}$ average distance, $\epsilon$ entropy, and $h$ level of hierarchy).

We show in Fig.~\ref{fig:emissions-pc1} the value of $\sum_c \tilde{g}_c$ as a function of the first morphological principal component, for extreme values of $\gamma$ and $d_0$. There seems to exist an optimal intermediate value for PC1 regarding the minimization of normalized indicator for emissions only. This would correspond to an intermediate level of monocentricity, meaning that urban areas which are too polycentric and spread would emit more, but also areas that are too much monocentric. This behavior does not occur for long-range $d_0 = 100$km and low-hierarchy $\gamma=0.5$ interactions. The mostly monocentric but emitting configurations capture most of population (given by the level of color), whereas the intermediate configuration capture around half of the population, what means that these low-emissions potential urban regions can cover a significant part of European population.

However, when considering both emissions and economic indicators, urban form then acts as a compromise variable. We show in Fig.~\ref{fig:paretos-relative} the point clouds of $\sum_c g_c$ against $\sum_c e_c$, which produce clear Pareto fronts, which shape varies with $\gamma$ and $d_0$. As the color level gives the value of PC1, we can see the points on the different fronts with very different morphological properties. In some case, highly monocentric areas (yellow points) can be a good compromise, whereas the intermediate optimal for emissions shown before may yield highly inefficient areas (dominated green points). For example, considering the fronts for $\gamma = 2$ which have both very similar shape, the points with the lowest emissions are on the top-left of the front and correspond to the optimal unveiled in Fig.~\ref{fig:emissions-pc1}. These have however a very low economic efficiency (high inefficiency) and small improvements can be done with the points below, before switching to a totally different urban form with a high value of PC1 (yellow points, highly monocentric). Increasing more the economic efficiency is then at the price of much more emissions, with more polycentric areas. This analysis therefore unveils morphological trade-offs, confirming that there is no optimal urban form, but different compromises regarding the conflicting sustainability indicators.

\section{Discussion}

\subsection{Developments}

Further work may consist in the use of calibration heuristics to find in a more robust way optimal parameter values. The OpenMOLE model exploration platform provides a transparent access to genetic algorithms for multi-objective optimization \citep{reuillon2013openmole}. The use of such calibration algorithms would allow to unveil the effective form of Pareto fronts, that we may have missed here through the grid sampling.

An other development would consist in extrapolating transportation flows with a spatially explicit gravity and transportation flow model as a kind of simplified four step model \citep{mcnally2000four}. It could then be adjusted on actual transportation flows emissions database which are also available in the Edgar database. The corresponding gravity parameters could then be used within the economic and emissions potentials, and the sustainability patterns produced compared with the hypothetical ones we produced here.

Finally, an important development would imply crossing our endogenous definitions of urban regions with socio-economic databases, and compute indicators implied in other dimensions of sustainability, for example related to socio-economic inequalities, spatial distribution of accessibilities, or activities with different scaling exponents. This includes the mitigation of spatial inequalities and segregation \citep{tammaru2015multi}, which are an important dimension of sustainibility.

\subsection{Towards policy applications}

Our work suggests the possibility to design policies in terms of regional integration to increase the sustainability of mega-city regions. The way such results could actually be transferred to policy-making recommandations remains an open question, but Pareto-optimal configurations can be used for the planning of regional transportation networks for example, or to design policies for the distribution of subsidies. Indeed, privileging some infrastructure developments but also collaborations between urban centers can be seen as an aspect of a small scale planning, or territorial strategy. As we integrated potential flows that would result from such development, and consider their economic and emissions consequences, and did it in an endogenous way, we suggest that evidence-based strategies for territorial development at the European level could be inspired by this work. This would naturally imply a more thorough data integration, model calibration and operationalization.

\section{Conclusion}

In conclusion, our multilayer percolation approach captures in a way the multi-dimensionality of urban systems and a link between form and function in urban system. Its application to the issue of sustainable mega-city regions shows its potentialities. This work also illustrates the importance of following data-driven paradigms even when developing, as what is understood of the behavior of the heuristic is through its application to real data and issues.












%


\end{document}